\documentclass[10pt,compsoc]{IEEEtran}
\usepackage{graphicx}      
\usepackage[pdflang={en-US},pdftex]{hyperref}
\usepackage{color}
\PassOptionsToPackage{warn}{textcomp}  
\usepackage{xspace}
\usepackage{xcolor}
\usepackage{caption}
\usepackage{subfigure}
\usepackage[normalem]{ulem}
\usepackage{enumitem}

\ifCLASSOPTIONcompsoc
  \usepackage[nocompress]{cite}
\else
  \usepackage{cite}
\fi
\newenvironment{tight_itemize}{\begin{itemize}[leftmargin=10pt] \itemsep
  1pt}{\end{itemize}}
  \setlist[itemize]{noitemsep, topsep=0pt}
\newcommand{\stitle}[1]{\noindent\textbf{#1}}

\newcommand{\cut}[1]{}
\newcommand{\change}[1]{#1}

\newcommand{\visrec}{VisRec\xspace}
\newcommand{\frontier}{\textit{Frontier}\xspace}
\newcommand{\baseline}{\textit{Baseline}\xspace}

\newcommand{\difference}{\texttt{Difference}\xspace}
\newcommand{\similarity}{\texttt{Similarity}\xspace}
\newcommand{\filter}{\texttt{Filter}\xspace}
\newcommand{\enhance}{\texttt{Enhance}\xspace}
\newcommand{\pivot}{\texttt{Pivot}\xspace}
\newcommand{\generalize}{\texttt{Generalize}\xspace}
\newcommand{\correlation}{\texttt{Correlation}\xspace}
\newcommand{\distribution}{\texttt{Distribution}\xspace}

\begin{document}
\title{{\LARGE Deconstructing Categorization in Visualization Recommendation:\\A Taxonomy and Comparative Study}}
\author{Doris Jung-Lin Lee, Vidya Setlur, Melanie Tory, Karrie Karahalios, Aditya Parameswaran
\IEEEcompsocitemizethanks{\IEEEcompsocthanksitem D. Lee and A. Parameswaran are with the University of California, Berkeley. V. Setlur and M. Tory are with Tableau Research. K. Karahalios is with the University of Illinois, Urbana-Champaign.
\IEEEcompsocthanksitem This work has been submitted to the IEEE for possible publication. Copyright may be transferred without notice, after which this version may no longer be accessible.}}
 
\IEEEpubid{0000--0000/00\$00.00~\copyright~2021 IEEE}


\IEEEtitleabstractindextext{%
\begin{abstract}
  Visualization recommendation (\visrec) systems provide users with suggestions for potentially interesting and useful next steps during exploratory data analysis. These recommendations are typically organized into categories based on their analytical actions, i.e., operations employed to transition from the current exploration state to a recommended visualization. However, despite \change{the emergence of a plethora of \visrec systems in recent work, the utility of the categories employed by these systems in analytical workflows has not been systematically investigated}.
Our paper explores the efficacy of recommendation categories by formalizing a taxonomy of common categories and developing a system, \frontier, that implements these categories. Using \frontier, we evaluate workflow strategies adopted by users and how categories influence those strategies. Participants found recommendations that add attributes to enhance the current visualization and recommendations that filter to sub-populations to be \change{comparatively} most useful during data exploration. 
Our findings pave the way for next-generation \visrec systems that are adaptive and personalized via carefully chosen, effective recommendation categories. 
%
\end{abstract}

}

\maketitle

\IEEEraisesectionheading{\section{Introduction}\label{sec:introduction}}
\begin{figure*}[h!]
   \centering
    \includegraphics[width=0.9\textwidth]{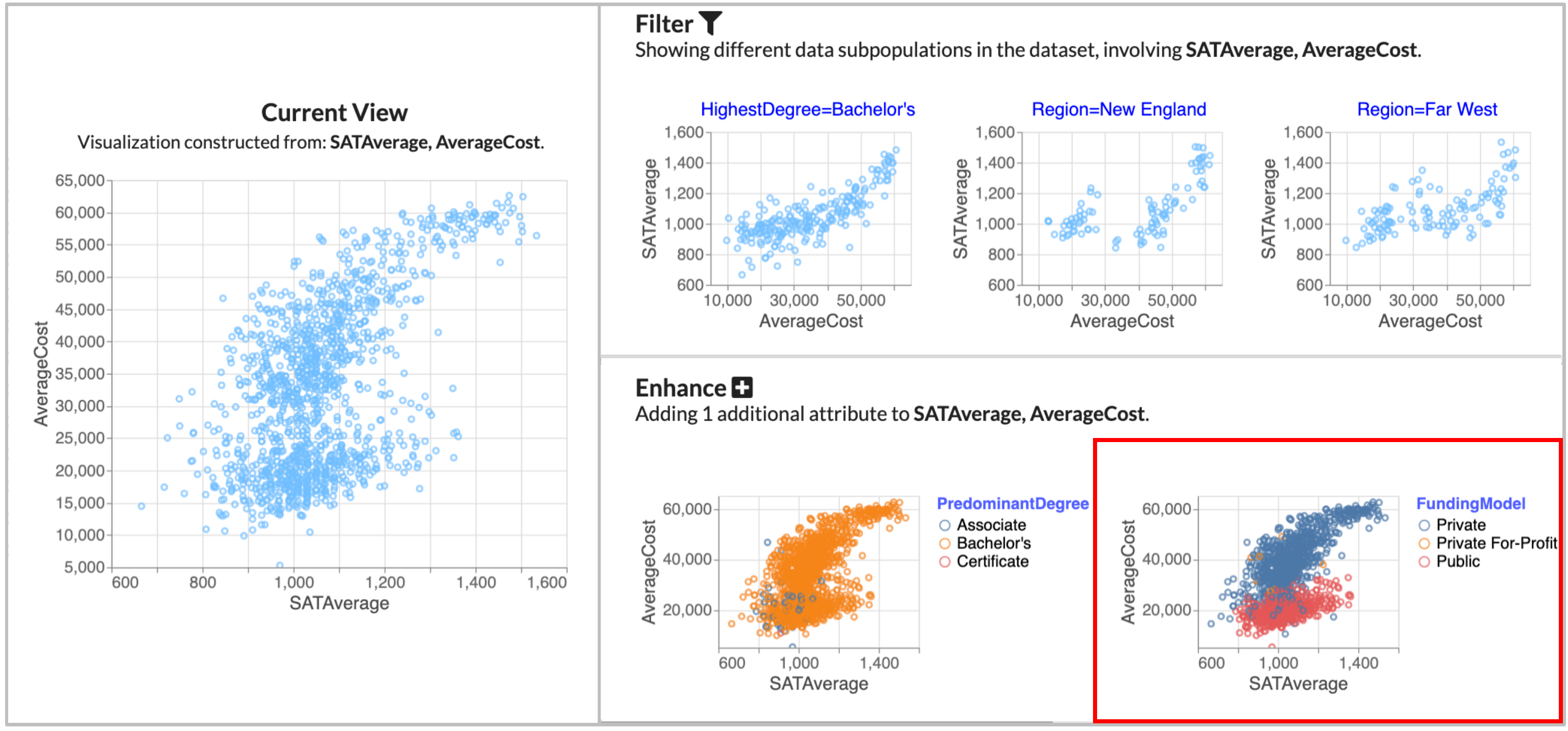}
    \caption{A screenshot of \frontier with a dataset containing college information. 
      Starting from the Current View displaying a scatterplot 
      of AverageCost versus SATAverage on the left, the
      user finds an interesting visualization recommended through
      the \enhance category highlighting the two 
      distinct clusters for Private and Public FundingModels (shown with a red border).
      This recommendation is generated from the Current View,
      further ``enhanced'' by adding FundingModel to the color channel.}
   \label{fig:teaser}
\end{figure*}
\par Exploratory visual analysis is an iterative process of asking and answering questions about data through visualizations, where new questions often arise from unexpected observations. Challenges arise when the current analysis path does not yield interesting observations; this common pain point can cause users to feel stuck or overwhelmed, unsure of what question to ask next~\cite{Grammel2010,zenvisage}. Visualization recommendation (\visrec) systems guide users along their exploration journey by suggesting effective visual encodings~\cite{Mackinlay2007,Wongsuphasawat2016,Wongsuphasawat2017} or potentially interesting visualizations~\cite{datasite,hu2018dive,Vartak2015,wills2010autovis,demiralp2017foresight,seo2005rank}.

\par Recommendations are often organized into categories based on the {\em analytical actions} they embody. {\em Analytical actions} can be thought of as transitions between visualization states, corresponding to the operations performed to generate recommendations given the current visualization state. Example categories include \filter, displaying recommendations of sub-populations of the data, derived by adding filters to the current visualization, and \enhance, displaying recommendations of an additional attribute added to the current visualization. Figure~\ref{fig:teaser} illustrates how recommendation categories can support the analysis of a college dataset. A scatterplot of \texttt{SATAverage} and \texttt{AverageCost} can be {\em filtered} to specific \texttt{HighestDegrees} or \texttt{Regions} (right side, top) or {\em enhanced} by adding the \texttt{FundingModel} attribute to the color channel (right side, bottom). Users can explore their data via moves in the visualization space, selecting visualizations based on analytical actions that represent potential ``next steps'' in their analysis.  

Most \visrec systems display a small set of bespoke analytical-action-based recommendation categories without a clear understanding of why the set was selected. This limited selection of categories in existing systems stems from challenges in both {\em development} and {\em evaluation}. From an evaluation standpoint, determining the value of a given recommendation for a specific user goal is, in general, a challenge in recommender system design~\cite{McNee2006}, but doing so for visual analysis tools is even harder. Unlike web search, where the typical goal is to find a single item (e.g. a movie to watch), insights in visual analytics often arise from multiple visualizations. This design is further complicated by the variety of user goals, ranging from specific inquiries to more complex open-ended objectives such as understanding relationships across attributes~\cite{munzner2015visualization}. From a development standpoint, there is an interface design and performance cost for dynamically computing large numbers of recommendations~\cite{Tunkelang:2009}. As a result, most systems rely on a small set of fixed recommendation categories.

While prior work has certainly demonstrated the benefits of \visrec in supporting exploratory analysis~\cite{datasite,Wongsuphasawat2016,Wongsuphasawat2017}, the design space of \visrec categories has not been thoroughly explored and evaluated. With new \visrec systems being introduced time and again in the visualization and HCI literature~\cite{lee2020medium}, there is a pressing need to take a step back to organize and make sense of the design space of analytical-action-based categories in \visrec, and to further evaluate the benefits of various types of recommendation categories as a whole, and relative to each other. \change{This evaluation is crucial for distilling design guidelines for next-generation \visrec systems and for enabling past, present, and future \visrec systems to be understood and compared in the context of an organization.}

In this paper, we deconstruct categorization in \visrec systems by comparing and evaluating the value of different recommendation categories in visual analytic workflows. We further investigate how analysis strategies are influenced by employing recommendation categories, as well as the efficacy of various recommendation categories for different task and dataset characteristics. While recommendation categories such as \filter and \enhance are in fact present in prior systems\cite{zenvisage,Lee:2019,van2013small,Wongsuphasawat2016,Wongsuphasawat2017,hu2018dive}, there has been no systematic organization or comparison of recommendation categories and their underlying analytical actions. Another challenge is that no existing \visrec system comprehensively implements the space of possible categories to compare their effects on analytical workflows. This crucial gap in existing literature motivated our system, \frontier \footnote{The name \frontier is inspired by how the application of analytical actions offer next steps along potential exploration paths---enabling an explorer to expand the \emph{frontier} of discovery. In the context of graph search algorithms, the term \emph{frontier} refers to the nodes that lie between what has been discovered and those as yet undiscovered~\cite{Cormen:2009}.}. We developed \frontier as an apparatus for investigating the merits and pitfalls of various recommendation categories in a single system.
\\\noindent Our contributions are summarized as follows:
\begin{tight_itemize}
\item We present a taxonomy of common analytical action-based recommendation categories employed in visual analysis, synthesizing existing literature from \visrec and online analytical processing (OLAP). The taxonomy enables us to map out the design space of existing \visrec systems as well as future ones. \change{(Section~\ref{sec:taxonomy})}
\item We develop a design probe, \frontier, implementing ten recommendation categories from the taxonomy to explore the usage and impact of these categories in a visual analysis workflow. \change{(Section~\ref{sec:system})}
\item  We present a mixed-methods user study to understand how recommendation categories support visual analysis and the relative efficacy of various recommendation categories. \change{(Section~\ref{sec:method},~\ref{sec:analysis})}
\end{tight_itemize}
\change{As part of this work, one of our goals was to take stock of and systematize research in} the rapidly-evolving area of \visrec systems. Our findings validate prior conjectures about the value of categorizing recommendations~\cite{Wongsuphasawat2016,demiralp2017foresight} and further reveal how there are substantial differences in the usage of various categories. For instance, participants indicated that recommendation categories \enhance (adding one attribute) and \filter (displaying data sub-populations) were most useful, while \pivot (swapping an attribute) was one of the least useful. Such findings point to the importance of comparative evaluation across categories and guidelines for improving category design in future \visrec systems presented in Section~\ref{sec:implication}.

\section{Related Work}
This work is draws from work on \visrec systems, faceted search interfaces, and traditional recommender systems. 

\subsection{Visualization recommendation systems}
Manual visualization specification tools~\cite{stolte2002polaris,Satyanarayan} require the user to specify the exact data attributes\change{, data subset of interest,} and the visual encodings for creating a visualization. This process is often tedious and overwhelming during the early, exploratory stages of analysis, especially for users without visualization design \change{or data exploration experience}~\cite{zenvisage,Wongsuphasawat2017,hu2018dive}. To alleviate this issue, \visrec systems suggest visualizations to assist users \change{with} visual analysis.

\par \visrec systems can be classified based on whether they suggest visual encodings (i.e., encoding recommenders) or aspects of the data to visualize (i.e., data-based recommenders)~\cite{wongsuphasawat2016towards}. The earliest \visrec systems focused on recommending visual encodings, assuming that the data attributes were already identified by the user~\cite{Mackinlay1986,Mackinlay2007}. Subsequent work automatically recommended graphical encodings based on perceptual effectiveness~\cite{moritz2019formalizing,Kim2017}. \change{Our focus in this work is on data-based recommenders, as in many recent papers ~\cite{wills2010autovis,Vartak2015,seo2005rank,wilkinson2005graph,demiralp2017foresight,Siddiqui2017,Lee:2019}, that suggest} interesting visualizations based on statistical properties of the data. To narrow the design space of recommendation categories, we leverage visualization best practices for determining visual encodings. \change{Throughout the rest of this paper, we use the term \visrec system to refer to those that employ data-based recommendations.}

\par Some \visrec systems are entirely automatic~\cite{wills2010autovis,Anand2015}, whereas other, more recent, mixed-initiative systems support some user interaction to guide the recommendations~\cite{demiralp2017foresight,Siddiqui2017,Wongsuphasawat2017}. Mixed-initiative \visrec systems combine manual specification with recommendations~\cite{hu2018dive,Key2012,yalccin2018keshif,van2013small,law2019duet,googleexplore, explorepatent,Lin2020}. For instance, Voyager~\cite{Wongsuphasawat2016,Wongsuphasawat2017,wongsuphasawat2016towards} suggests visualizations based on user-selected fields and wildcards to iterate through possible data attributes or encodings. Many of these systems organize the resulting recommendations into categories based on their analytical actions~\cite{Wongsuphasawat2016, Wongsuphasawat2017, hu2018dive, demiralp2017foresight,yalccin2018keshif}. For example, DIVE~\cite{hu2018dive} and Voyager display a set of univariate distributions to help users get an overview of the distributions that exist in the data as well as recommendations that add an additional attribute to the user-selected visualizations. On the other hand, Zenvisage~\cite{Siddiqui2017} searches for visualizations based on their similarity with a user-specified pattern over a space of possible filter values. While intuitively valuable, it remains unclear what the effects of different categories are on users and which categories are most helpful. Our work specifically investigates analytical action-based recommendation categories and how they impact mixed-initiative visual analytics workflows. 

\subsection{Faceted search interfaces}
Facets help users quickly refine their search options by applying multiple filters based on a categorized classification of the information elements in web and product search~\cite{HearstFacettedBrowsing,Yee:2003} and text corpora querying~\cite{Cao2010,Collins2009,Dork2012} interfaces. Faceted search interfaces bear similarities to \visrec systems in that both support progressive disclosure and incremental construction of queries where users can formulate the equivalent of a sophisticated data query through a series of small, exploratory steps~\cite{Yee:2003,English2002}. One key problem in building a faceted search interface is selecting the facets. Some systems simply present the first few alphabetized facets~\cite{Hearst:2002}; others present a subset of facets ranked by frequency of use~\cite{ebay,amazon,netflix}. Prior \visrec systems have drawn an analogy between faceted browsing and exploratory visual analysis~\cite{Wongsuphasawat2016,datasite}. Similarly, our work draws inspiration from principles of facets as recommendation categories to address the complexity of analytical tasks~\cite{munzner2015visualization}. While facets are constructed based on aspects of items in the corpus (e.g., size, brand for products; topics for articles), in \visrec, categories can be organized based on their relationship with the current visualization or visual characteristics as discussed in more detail in Section~\ref{sec:taxonomy}.

\begin{table}[]
	\centering
	\begin{tabular}{l|l}
		Category               & Related Work                                                                                                            \\ \hline
		Distribution           & \cite{hu2018dive,Wongsuphasawat2016,Wongsuphasawat2017,Sarvghad2017,wills2010autovis,demiralp2017foresight,seo2005rank} \\
		Correlation            & \cite{hu2018dive,Wongsuphasawat2017,Srinivasan2019,wills2010autovis,demiralp2017foresight,Dang:2014,datasite}           \\
		Enhance                & \cite{hu2018dive,Wongsuphasawat2016,Wongsuphasawat2017,van2013small,Lin2020}                                            \\
		Generalize (attribute) & \cite{hu2018dive,Lin2020,seo2005rank}                                                                                   \\
		Generalize (filter)    & \cite{hu2018dive}                                                                                                       \\
		Pivot                  & \cite{hu2018dive,van2013small,Lin2020}                                                                                  \\
		Filter (add)           & \cite{van2013small,Anand2015,Lee:2019}                                                                                  \\
		Filter (swap)          & \cite{zenvisage}                                                                                                        \\
		Difference             & \cite{law2019duet,Vartak2015,Anand2015}                                                                                 \\
		Similarity             & \cite{Law2020,zenvisage}                                                                                               
	\end{tabular}
	\caption{Survey of recommendation categories from 20 \visrec systems. We included systems that \change{recommend} visualization(s) based on the result of some analytical action.}
	\label{tab:landscape}
\end{table}

\vspace{-5pt}
\subsection{Recommender Systems}
Unlike search systems that aim to maximize the relevance of retrieved items, the goal of a recommender system is to help users discover items of interest. As a result, an ideal recommender system must strike the right balance between suggesting items that are relevant as measured by recommender accuracy, and \change{those} that are diverse, \change{and are therefore} surprising and unexpected.
\par This diversity-accuracy tradeoff has been well-studied in the recommender system literature and has led to metrics beyond accuracy such as serendipity, novelty, coverage, and diversity~\cite{Vargas2011,Kaminskas2016,Clarke2008}. The tradeoff manifests itself in visual data exploration as well, where users often want to discover non-obvious, unexpected insights, but still want to see visualizations relevant to the attributes or values that they are interested in. The context-dependent actions in the taxonomy described next, are examples of visualization recommendations relevant to the user's context; context-independent actions can reveal more surprising aspects about the data. Akin to diversification in traditional recommender systems, our paper highlights the importance of surfacing a relevant, yet diverse set of potential next steps that satisfies the user's information needs in a visual analytic workflow.

\section{Taxonomy of Recommendation Categories \label{sec:taxonomy}}
Analytical actions correspond to transitions through the visualization space to generate categories of recommendations given a user's current visualization state. While various taxonomies~\cite{Munzner2014,Brehmer2013,Amar2005} exist for describing the types of tasks (or \emph{actions}) employed during visual analysis, \change{as stated in Law et al.~\cite{Law2020}, we are not aware of any taxonomy that encompasses} the types of data-based visualization recommendations that can be generated via analytical actions \change{and are reflected in present-day \visrec system}. This section defines \change{such a taxonomy, providing} a common vocabulary for the organizing principles behind recommendation categories. The taxonomy arose through a systematic review of 20 \visrec systems detailed in the supplementary material. 

\begin{figure}
  \centering   
  \vspace*{-10pt}
  \includegraphics[width=\linewidth]{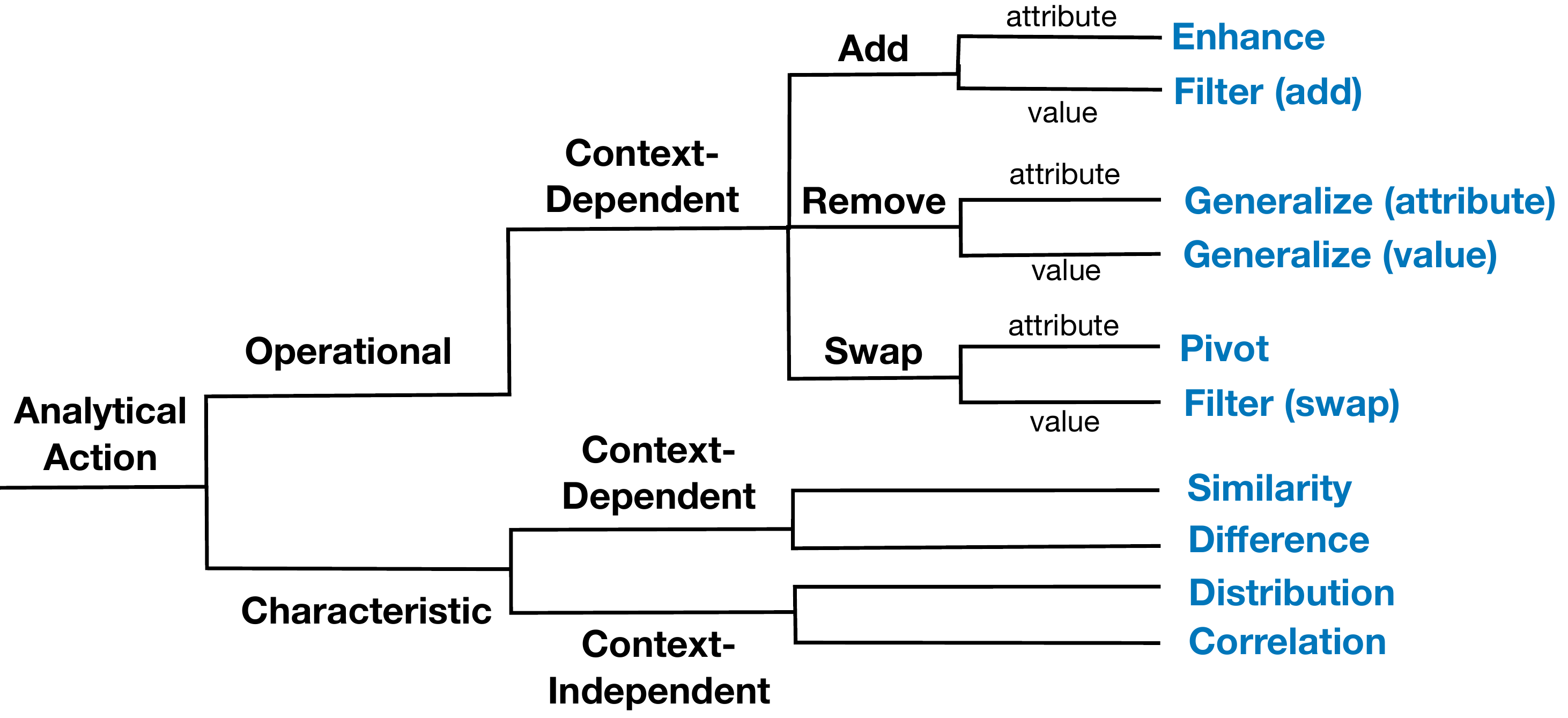}
  \caption{A taxonomy of common analytical actions used in recommending visualizations for visual analysis. The analytical actions are indicated in blue.}
  \label{fig:taxonomy}
  \vspace*{-10pt}
\end{figure}

\par Through open and axial coding~\cite{groundedTheory}, we uncovered ten major types of analytical actions used to generate and group visualization recommendations in existing systems, summarized in Table~\ref{tab:landscape}. To keep the design space of recommendation categories tractable, we focused on data-based recommendations driven by the operational and characteristic transitions in the visualization design space. We codified these actions into a taxonomy as seen in Figure~\ref{fig:taxonomy}. The table is not intended to capture a comprehensive set of all analytical action types, but rather to synthesize the most common ones used to categorize recommendations so that we can explore the interaction design space more deeply.
\par At its highest level, the taxonomy defines two main categories: \textit{operational} and \textit{characteristic}. The \emph{operational} category describes analytical actions that navigate users through the visualization space via operations such as add, remove, and swap. The \emph{characteristic} category describes actions that reveal certain characteristic patterns in the data, such as skewness and correlation. The taxonomy is further broken down into \textit{context-dependent} and \textit{context-independent} categories. Actions are \emph{context-dependent} if they depend on the \emph{current view} or visualization (i.e., the selected attributes, values, and visual encodings); they are \emph{context-independent} if they do not depend on the \emph{current view}.
\par \change{Note that the operational actions described above} overlaps with some of the categories in the chart transition model in GraphScape~\cite{Kim2017}. However, GraphScape \change{provides} a chart transition model that describes visualization edits, whereas our taxonomy describes actions in a \visrec context \change{drawn from existing \visrec systems}. Since our focus is less on encoding-based recommendations, we do not consider \change{such} aspects of the GraphScape taxonomy, such as Scale and Mark. Likewise, GraphScape does not include characteristic actions described in this taxonomy. 
\vspace*{-5pt}
\subsection{Operational Analytical Actions}
Operational actions apply data-oriented operations that transition the current view to a related, neighboring part of the visualization space. By definition, analytical actions that are operational must also be \emph{context-dependent} as they operate on the current view. As seen in Figure \ref{fig:taxonomy}, there are three broad categories of operational actions based on whether an attribute or value is added, removed, or swapped, leading to six ($3 \times 2$) individual categories. 

\par The example in Figure~\ref{fig:hierarchy} demonstrates how operational actions can be thought of as moving along different paths in the \emph{attribute} or \emph{value} hierarchy. Every node in the attribute or value hierarchy represents a set of selected attributes or values. A user's current visualization is composed of their position on the attribute hierarchy (i.e., the space of all attribute combinations) and their position on the value hierarchy (i.e., the space of all filter value combinations). Movements through these hierarchies defines the set of possible operational actions. This conceptual model formalizes the space of possible visualizations that are one move away from the current visualization. \change{Our model draws on Online Analytical Processing (OLAP), a sub-field of data management that targets analytical querying of multi-dimensional data. However, unlike OLAP, which only considers the value hierarchy~\cite{Gray1997}, we introduced the analogous attribute hierarchy to help capture common operations in visual analytics.} 
Here are the six operational analytical action categories:

\begin{figure*}[t]
  \vspace{-5pt}
  \centering
  \includegraphics[width=0.9\linewidth]{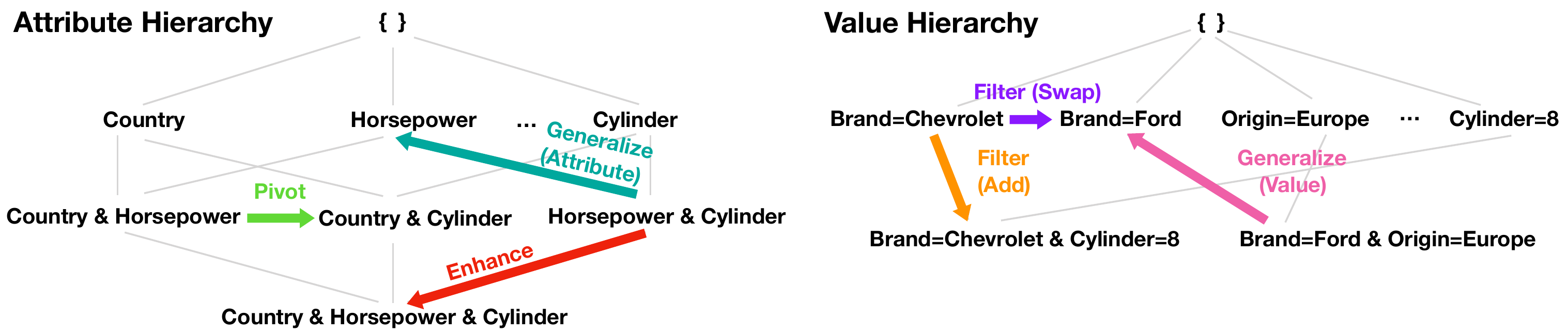}
  \vspace{-5pt}
  \caption{Operational actions represent transitions through the attribute and value hierarchies.}
  \label{fig:hierarchy}
  \vspace{-5pt} 
\end{figure*}

\begin{tight_itemize}
\item \stitle{\enhance:} adds an additional attribute to the current view. If the user selects attributes $A$ and $B$, \enhance displays visualizations involving attributes $A$, $B$, and $C$. This action corresponds to moving down the attribute hierarchy (Figure~\ref{fig:hierarchy} red).
\item \stitle{\texttt{Filter (add)}:} adds an additional filter to the current view. If the user selects attributes $A$ and $B$, \texttt{Filter (add)} displays visualizations involving $A$, $B$, and a filter $F$. In OLAP~\cite{Gray1997}, this is known as a \emph{drill-down} on the value hierarchy (Figure~\ref{fig:hierarchy} orange). 
\item \stitle{\texttt{Filter (swap)}:} switches out the filter value to a different value, while keeping the filter attribute fixed. If the user selects attributes $A$ and filter $F=V$, \texttt{Filter (swap)} displays visualizations involving $A$ and an alternative filter $F=V'$. This action corresponds to moving horizontally across the value hierarchy to a node with the same filter attribute (Figure~\ref{fig:hierarchy} purple).
\item \stitle{\texttt{Generalize (attribute)}:} removes one attribute from the current view to display the more general trend. If the user selects attributes $A$ and $B$, visualizations involving either $A$ or $B$ are displayed. This action corresponds to moving up the attribute hierarchy (Figure~\ref{fig:hierarchy} turquoise).
\item \stitle{\texttt{Generalize (value)}:} removes one filter from the current view to display the more general trend. If the user selects attributes $A$ and a filter $F$, visualizations involving only $A$ are displayed. In OLAP, the removal of a filter is known as a \emph{roll-up} on the value hierarchy (Figure~\ref{fig:hierarchy} pink).
\item \stitle{\pivot:}  displays visualizations that can be constructed if one of the attributes from the current view is replaced with another attribute. If the user selects attributes $A$ and $B$, \pivot displays visualizations involving either $A$ and another attribute $B'$, or $B$ and another attribute $A'$. This action corresponds to moving horizontally along the attribute hierarchy (Figure~\ref{fig:hierarchy} green). 
\end{tight_itemize}
\vspace*{-3pt}
\subsection{Characteristic Analytical Actions}
\vspace*{-2pt}
Characteristic analytical actions are designed to surface salient visual and statistical characteristics of the data, sorted based on an interestingess metric that is described in Section~\ref{sec:objective}. Characteristic actions that are \emph{context-independent} are designed with an overview intent, similar to ``breadth-first'' exploration strategies in web search~\cite{Tunkelang:2009} that are independent of the user's search query. We describe two types of independent actions that highlight patterns that may be of interest to the user:
  \begin{tight_itemize}
  \item \stitle{\correlation:} highlights bivariate relationships between quantitative fields in the data through scatterplots of different combinations of quantitative attributes.
  \item \stitle{\distribution:} displays the possible univariate distributions in the dataset, with \texttt{COUNT} as the default measure. The visualization can either be a histogram, bar chart, or line chart depending on the data type of the attribute.
  \end{tight_itemize}
Characteristic actions that are \emph{context-dependent} showcase salient visual characteristics based on the current view.
  \begin{tight_itemize}
    \item \stitle{\similarity/\difference:} highlights data patterns that are visually similar or different from the current view. 
    \end{tight_itemize}
\vspace{-5pt}

\subsection{Ranking Objectives\label{sec:objective}}
Within each analytical action category, visualizations
are often ranked using some interestingness objective. 
Given the different chart characteristics for different types of visualizations, the interestingness objective, even for a given action, may be different for visualization types. \change{For example, a user may be interested in the degree of correlation in a scatterplot, while they may be interested in differences between the bar values in a bar chart.} In this paper, we consider \change{commonly occurring} basic chart types, including bar charts, histograms, line charts, and scatterplots, \change{typically employed by existing \visrec systems}. \change{Even this set results} in a \change{considerable} number of choices
corresponding to every combination of 
action and visualization type. We identified a small number
of classes of objectives that have been used in prior work
for these combinations, which we catalog below.

For the {\em characteristic} actions, the objective typically captures the salient visual characteristics expressed by a visualization, such as the degree of correlation or skew. \change{Visualizations are ranked} from the \correlation action
based on monotonicity~\cite{wilkinson2005graph}, typically most to least correlated,
while those from the \distribution action \change{are ranked} from most to least skewed.
For the \similarity action, bar and line chart visualizations
are ranked based on similarity to the {\em current view},
computed via the Euclidean distance between the measure values of the visualizations~\cite{Siddiqui2017,zenvisage}.

For the {\em operational} actions,
the objective used is typically determined by 
the visualization type of the recommended visualizations. 
These objectives capture 
perceptual characteristics 
generally associated with 
something unexpected or insightful in the visualization, including: 
   \begin{tight_itemize}
    \item \textbf{Non-uniformity}: For \textit{bar/line charts and histograms without a filter}, visualizations are \change{ranked highly} if they are highly uneven, indicating the presence of outlying categories or shifts in distributions~\cite{demiralp2017foresight, datasite}. 
    \item \textbf{Deviation}: For \textit{bar/line charts and histograms with a filter}, the \change{ranking} is based on the deviation between the filtered and unfiltered (overall) distributions, based on the intuition that a visualization is potentially interesting if it differs greatly from some expected reference~\cite{Lee:2019,Vartak2015}.
    \item \textbf{Correlation}: For \textit{uncolored scatterplots}, a
    visualization is \change{ranked higher} if it displays a high degree of dependence between the two measures, as measured by mutual information~\cite{Murphy2012,kandel2012profiler} or Spearman's correlation~\cite{wilkinson2005graph}.
    \item \textbf{Separability:} For \textit{colored scatterplots}, a visualization is \change{ranked higher} if the colors for each category distinctly separate clusters of data points in the scatterplot~\cite{Sedlmair2012,datasite}.
  \end{tight_itemize}

\noindent Supplementary materials provide details as to how these objectives were implemented in our design probe, \frontier.

\section{The Frontier System\label{sec:system}}
\begin{figure}[h!]
  \vspace{-10pt}
    \includegraphics[width=\linewidth]{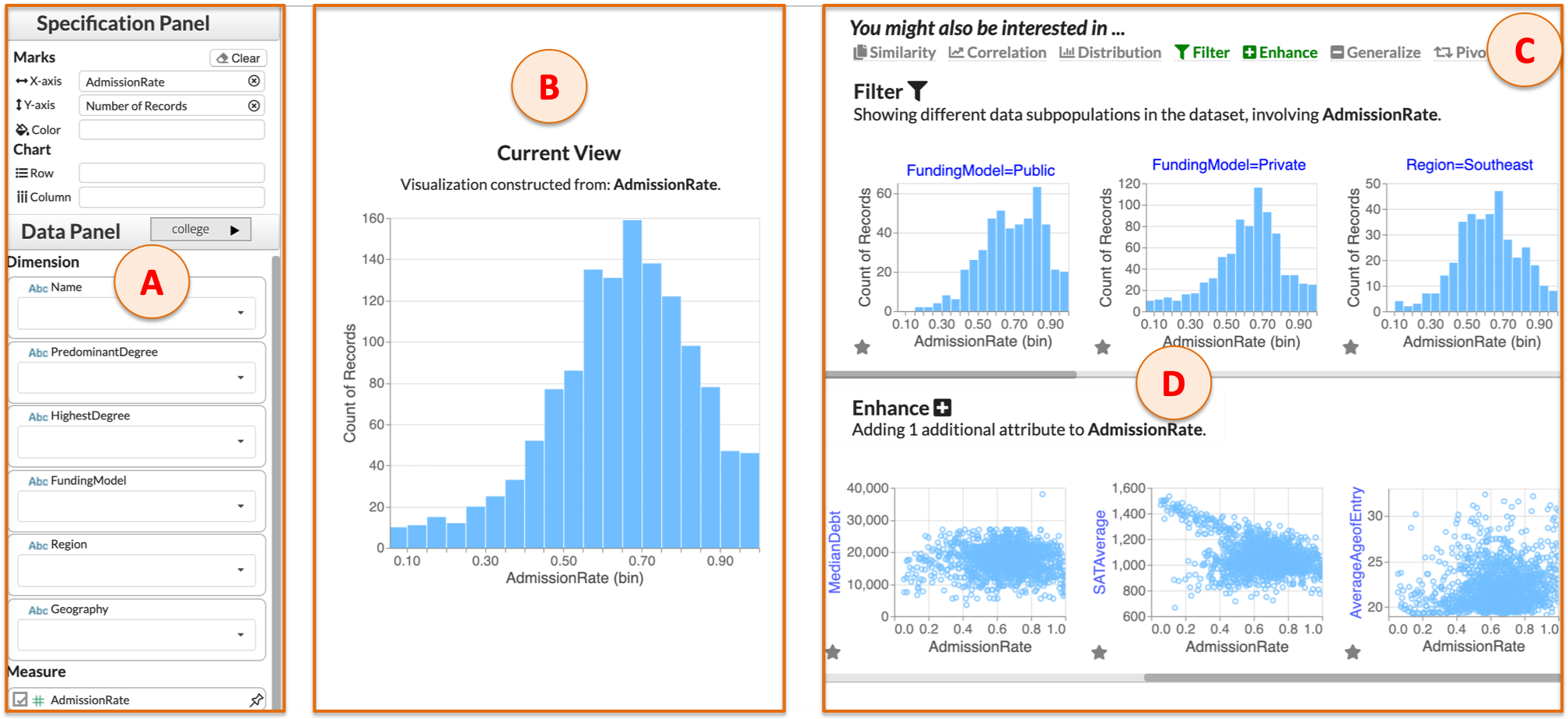} 
    \vspace{-5pt}
    \caption{\frontier consists of four areas: Control Panel (A), Current View (B), Category Menu (C), and Recommendations Panel (D).}
    \label{fig:interface}
    \vspace{-10pt}
\end{figure}

We introduce a system, \frontier, that provides visualization recommendations across multiple analytical action-based categories. \frontier is a design probe that enables us to systematically explore and compare these categories. For brevity, we refer to the analytical action-based categories displayed in \frontier simply as \emph{recommendation categories} henceforth. The \frontier interface is composed of four areas as illustrated in Figure~\ref{fig:interface}. Starting from the left (Figure~\ref{fig:interface}A), we have the Control Panel, a manual specification interface for specifying the visualization in the Current View (Figure~\ref{fig:interface}B). The Control Panel lists measures and dimensions and allows users to add or remove attributes and values. The Specification Panel (Figure~\ref{fig:interface}A top) allows users to fine-tune their visualization by arranging attributes across specific encoding channels. Users can toggle on and off specific recommendation categories. If a category is not applicable for the given Current View, the cursor icon changes to a forbidden sign upon hover in the Category Menu (Figure~\ref{fig:interface}C). The recommendations are displayed row-by-row on the right (Figure~\ref{fig:interface}D) analogous to faceted web search results to encourage browsing~\cite{Yee:2003}. 
We drew inspiration from existing \visrec interfaces~\cite{Wongsuphasawat2017,hu2018dive,demiralp2017foresight} and employed guidelines from mixed-initiative interfaces~\cite{Horvitz:1999,Nielsen1990} to balance interface usability with comprehensiveness in the display of recommendation categories. We iterated on the design with feedback from an interaction designer and made significant changes to the interface over a period of six months. 
\subsection{Design considerations}
The following design considerations emerged while iteratively designing \frontier:
\begin{tight_itemize}
\item C1: \emph{Concise and informative.} Recommendation categories should provide a \emph{manageable} set of options as ``next-steps'' in a user's analytical workflow. Users should never be shown an empty category nor should they be shown categories with overlapping recommendations.
\item C2: \emph{Coordinated and actionable.} Recommendation categories should be coordinated and consistent with other parts of the system, such as in the Category Menu and Current View. The user should be able to bring a recommended visualization into the Current View.
\item C3: \emph{Interpretable and visually discernible.} Recommendation categories should be self-explanatory and display visual indicators that convey their key characteristics or highlight how they differ from the Current View.
\end{tight_itemize}
\change{These requirements echo design considerations from prior work in mixed-initiative visual analytics systems~\cite{Wongsuphasawat2017,Srinivasan2019}.}
\subsection{System Overview}

\frontier is a web-based system with components described as follows. First, the \emph{Data Manager} loads the dataset and metadata (i.e., the data type, data model, and default aggregation) and computes statistics (i.e., cardinality, correlation, minimum, maximum). The \emph{Context Manager} maintains information about the attributes and values that the user has selected. 
Then the \emph{Category Manager} determines which recommendation categories to display for a given Current View and maintains a list of categories. Finally, each \emph{Category} contains information about specific recommendation categories, a sorted list of top-$k$ recommended visualizations, and their associated scores. Details of the system architecture and implementation along with source code can be found in the supplementary materials.
\vspace{-10pt}
\subsection{Category Generation and Visualization Interaction}
\vspace{-3pt}
\par To select a manageable set of recommendation categories (C1) to display to users, we designed the following workflow. These rules are similar to the ones adopted in Voyager~\cite{Wongsuphasawat2016} and DIVE~\cite{hu2018dive}, which first provide an overview via univariate distributions, followed by more relevant visualizations based on subsequent user selection.
\par
\begin{figure*}[h!]
  \begin{center}
  \vspace*{-5pt}
  \setlength{\fboxrule}{0.25pt}
  \setlength{\fboxsep}{0pt}
  \fbox{\subfigure[Correlation]{ \includegraphics[width=2.25in,height=0.7in]{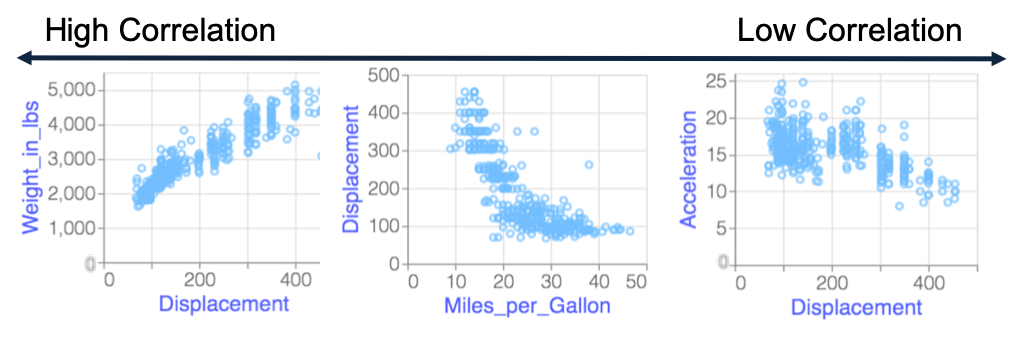}}}
    \hspace{0.05pt}
  \fbox{\subfigure[Distribution]{ \includegraphics[width=2.25in,height=0.7in]{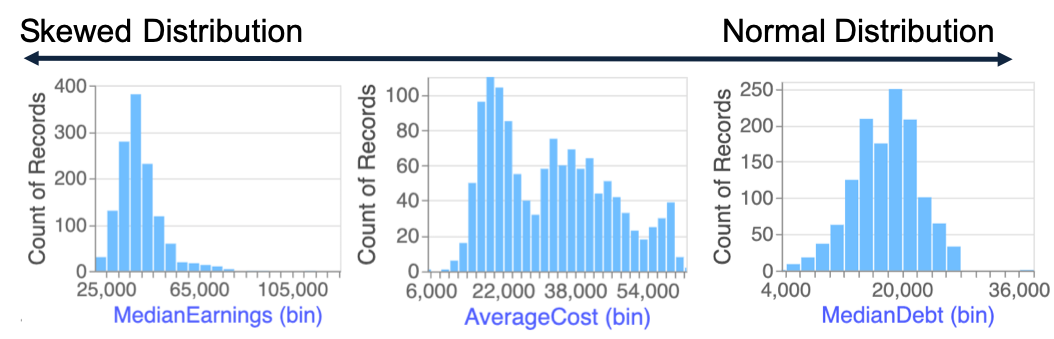}}}
    \hspace{0.05pt}
  \fbox{\subfigure[Generalize]{ \includegraphics[width=2.25in,height=0.7in]{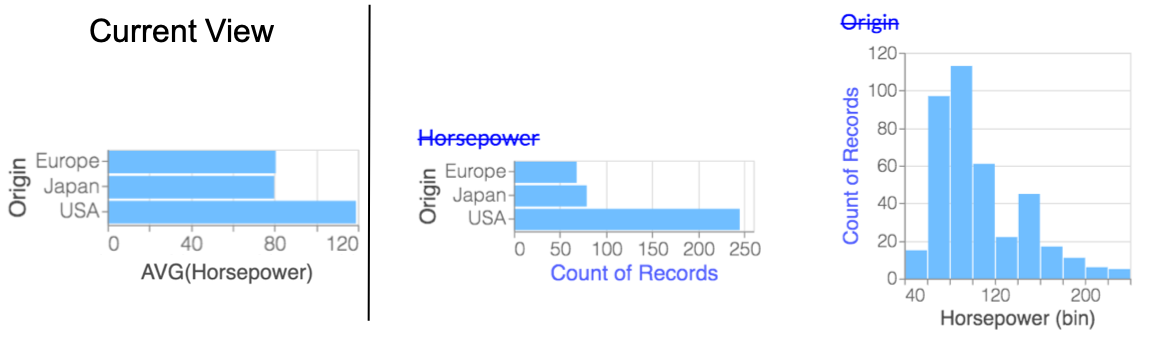}}}\\
  \par\vspace{.3mm}
  \fbox{\subfigure[Similarity]{ \includegraphics[width=3.5in]{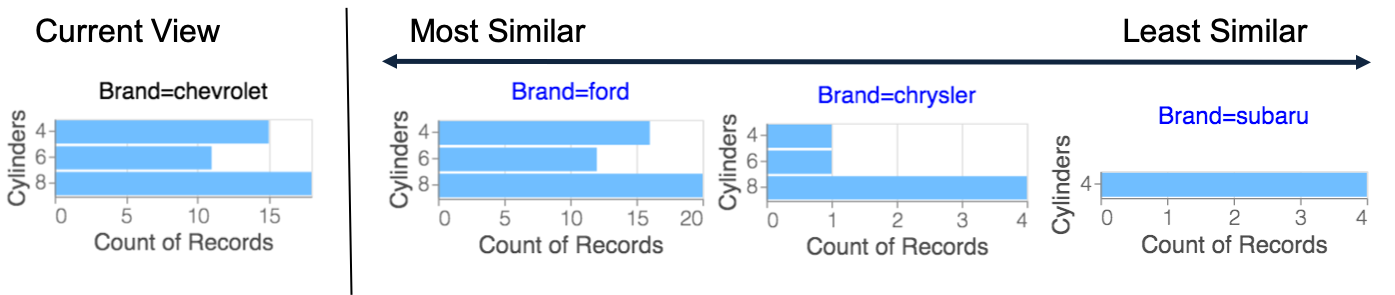}}}
    \hspace{0.05pt}
  \fbox{\subfigure[Pivot]{ \includegraphics[width=3.5in]{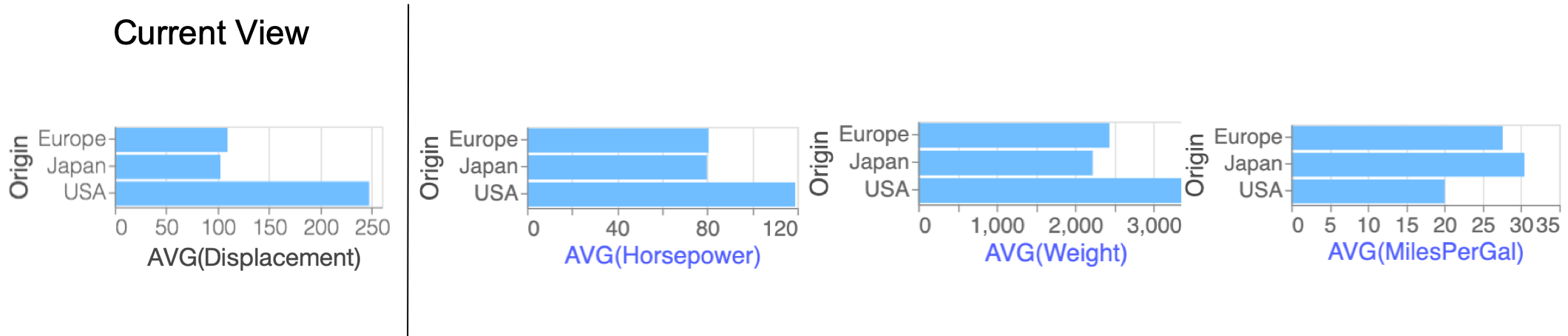}}}\\
  \par\vspace{.3mm}
  \fbox{\subfigure[Enhance]{ \includegraphics[width=3.5in]{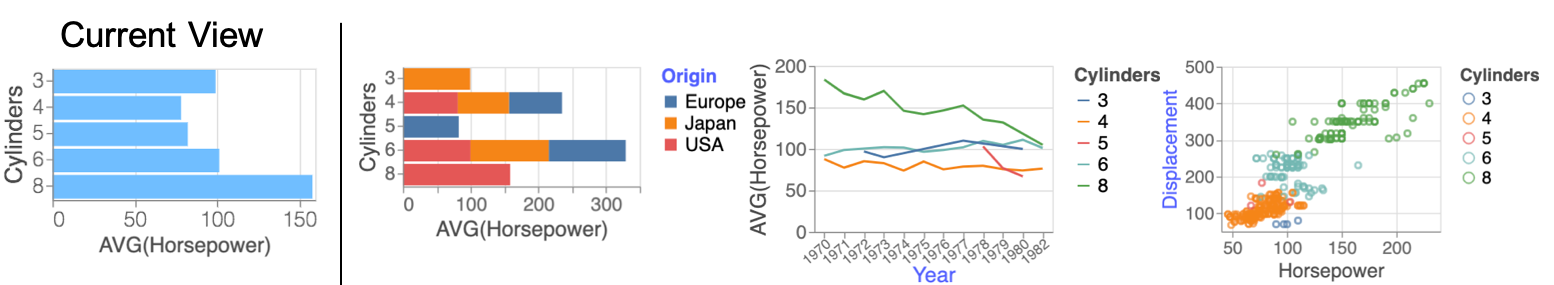}}}
     \hspace{0.05pt}
  \fbox{\subfigure[Filter]{ \includegraphics[width=3.5in]{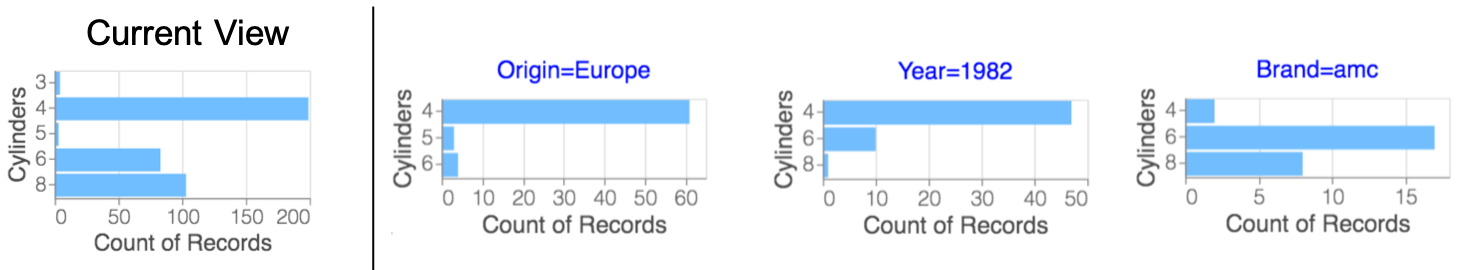}}}  
    \vspace*{-5pt}
      \caption{Examples of various recommendation categories implemented in \frontier. (A) \correlation generates scatterplots with bivariate relationships between quantitative fields ranging from high to low correlation. (B) \distribution shows the possible univariate distributions from the dataset ranging from skewed to normal distributions. In the following examples, the current view is shown on the left, with the corresponding recommendations shown on the right. (C) \generalize shows possible visualizations when one attribute or filter from the current view is removed (removed attributes shown with strikethroughs). (D) \similarity highlights data patterns ranging from most to least similar to the current view. (E) \pivot shows possible visualizations that can be constructed if one of the current attributes is changed to another (changed attributes shown in blue). (F) \enhance shows possible visualizations when an additional attribute is added to the current view (additional attributes shown in blue). (G) \filter displays the data subsets that can be constructed from the current view when a filter is applied.}
         \label{fig:facettypes}
    \vspace*{-5pt}
  \end{center}
  \end{figure*}

 At the start of the analysis, when no attributes are selected, \change{the \correlation and \distribution actions display} univariate and bivariate visualizations, enabling users to get an overview of the dataset (Figure~\ref{fig:facettypes}A, ~\ref{fig:facettypes}B). Operational categories evolve based on the Current View and come into play once any attribute is selected. To avoid redundancy across categories (C1), \frontier only shows context-independent categories when there are no attributes selected. For example, when a user selects a single quantitative attribute, \enhance (Figure~\ref{fig:facettypes}F) generates a collection of scatterplots, similar to what is shown for \correlation. Similarly, when a categorical attribute is in the Current View, only \pivot recommendations (Figure~\ref{fig:facettypes}E) are displayed, to avoid operating over the same collection \change{of visualizations as the ones in} \distribution. 

To make the recommendation categories more succinct and manageable (C1), from Table~\ref{tab:landscape}, \frontier consolidates filter (add) and filter (swap) to give \filter, generalize (attribute) and generalize (value) to give \generalize, similarity and difference to give \similarity. \frontier's \filter adds an additional filter to the Current View when there is no filter in the Current View (Figure~\ref{fig:facettypes}G). When a filter is in place, \filter keeps the specified filter attribute, while swapping out one of the attribute values to showcase alternative data subsets for comparison. \change{Note that} any applied filter is always retained in all actions except in \filter. In \generalize, we display all possible visualizations by removing either one filter or one attribute that is in the Current View (Figure~\ref{fig:facettypes}C). In \similarity, visualizations that look most similar to the Current View are ranked highest, but users can reverse the sort order to look at the most dissimilar visualizations (Figure~\ref{fig:facettypes}D). 

Users can double click any recommendation to bring the visualization into the Current View; this sets elements in the Control Panel to be consistent with that of the selected visualization (C2). We display the axis label of any element that differs between the Current View and the recommendation in blue, to ease comparison and highlight differences (C3). 
\vspace{-5pt}

\section{Study Design\label{sec:method}}

We conducted a mixed-methods study
to explore how various recommendation categories
impact visual analysis workflows.
Our primary goal was to study the relative
usefulness of various categories
as well as how categorization influences
the analysis workflow in general.
To study these goals, our design probe, \frontier,
implements the categories described in the previous section.
Further, to understand the effect of categorization in mixed-initiative \visrec workflows, we included
a mixed-initiative \visrec baseline, mirrored off
of \frontier, that featured the same recommendations, 
but without any categorization. 
We describe this baseline later on.
We opted against comparing with manual specification tools
without recommendations, given the general benefits
of \visrec in prior studies~\cite{datasite,Wongsuphasawat2017,Wongsuphasawat2016}.
We also opted against comparing with existing \visrec systems 
that implement a subset of categories, as this would not allow us to tease apart the impact of categorization on analytic workflows.


Overall, our exploratory study aimed to address the following research questions:
\begin{tight_itemize}
	\item RQ1: How do recommendation categories support and influence analytical workflows? What problem-solving and exploration strategies do users adopt when using recommendation categories in a mixed-initiative context? 
	\item RQ2: What are the differences in user behavior across recommendation categories? What is the value and impact of individual recommendation categories and how does this vary across tasks and datasets? 
\end{tight_itemize}
\subsection{Participants}
We recruited $24$ participants ($10$ female, $14$ male) from within a software company. Nine were experienced users of a popular, commercial charting tool, $13$ had limited proficiency, and two had no experience. In a between-subjects design, participants were randomly assigned to use \frontier or \baseline with either the College~\cite{college} or Olympic Medals~\cite{medals} dataset, with six participants per condition-dataset combination. Henceforth, we suffix .F or .B in the identifier to display whether the participant used \frontier or the \baseline condition.
\subsection{Tasks and Data}

There were two main parts to the study: closed-ended tasks and open-ended exploration. 
\subsubsection*{\textbf{Part 1: Closed-ended tasks}}
Closed-ended tasks were mainly intended to familiarize participants with the system while
also providing some consistent objectives for task comparison. Participants completed four closed-ended questions that included common visual analytic tasks, including: 
\begin{tight_itemize}
\item Q1 (\textsl{Correlate}): find other measures that are linearly correlated with a selected attribute.
\item Q2 (\textsl{Filter Compare}): compare bar charts across different data sub-populations. 
\item Q3 (\textsl{One v.s. All}): compare a filtered distribution with the overall distribution. 
\item Q4 (\textsl{Pattern}): compare the temporal trend across different measures.
\end{tight_itemize}
For each task, participants answered a multiple choice question on a paper worksheet. They were instructed to use \frontier to answer the question, but were not told how to do so. All participants used the Cars dataset~\cite{cars} for closed-ended tasks. This dataset was chosen because of its simple schema (five measures and five dimensions), clean insight patterns, and because it is commonly used for demonstrating visualization systems~\cite{Satyanarayan,Wongsuphasawat2017,datasite}, thus enabling comparisons.

\subsubsection*{\textbf{Part 2: Open-ended exploration}}
Following the closed-ended tasks, participants completed an open-ended exploration task. This task enabled us to observe how people would choose to use (or not use) the recommendations in a natural analysis flow.
Participants explored either the College or Olympic Medals dataset. Instructions were: ``\textit{We'd like you to explore this data to look for interesting insights. As you work, please let us know what questions or hypotheses you're trying to answer as well as any insights that you are learning about the data.}'' Participants were instructed to talk aloud and star recommendations they found useful.

The two datasets for open-ended exploration were chosen due to their real-world and accessible nature. We chose datasets with different characteristics, enabling us to study a wider range of analytical inquiries: the College dataset has ten measures and six dimensions with low to medium cardinality, while the Medals dataset contains only three measures and twelve dimensions with medium to high cardinality. 

\subsection{Apparatus}
In the \frontier condition, participants used the full version of \frontier with all of the recommendation categories. The ordering of recommendation categories within the interface was randomized for each user to minimize the preference for recommendations displayed at the top of the page. 

To study the impact of categorization as a whole, we introduced a \baseline condition. This condition displayed the same set of recommendations except that the recommendation categories were removed so that all the recommendations appeared in a single, grid layout\footnote{See screenshot in supplementary material. The \baseline interface has a similar layout to several existing \visrec systems~\cite{wills2010autovis,Key2012}}. The goal of this \baseline is to establish a vanilla \visrec system that {\em (i)} eliminates the effects of recommendation categories, while preserving certain characteristics for a controlled comparison in that {\em (ii)} it is mixed-initiative and {\em (iii)} displays the same overall set of recommendations. 

\par To understand this condition better, note that the organization of \visrec into categories is a result of both the labeling (i.e., interface elements such as dividers and textual descriptions of the categories) as well as the ranking within each category. To remove the effects of categorization {\em (i)}, we not only had to remove the category labels, but also had to shuffle the display order of the recommendations. 
In both conditions, participants can browse for more visualizations via horizontal scrolling (single scroll bar for the \baseline, one scroll bar per recommendation category for \frontier). To prevent preferential bias towards top-ranked visualizations, we further ensured that the exact same set of visualizations appeared with and without scrolling across both conditions.
\par While we acknowledge that our baseline is not perfect, we considered alternative designs, including a no-recommendations baseline or no baseline at all. However, a baseline that only removes the category labels, but does not alter the display order would only evaluate the effects of explicit category labels and is thus not a meaningfully different baseline. Since the goal of the study is not to demonstrate performance difference between \frontier and \baseline, we opted for a baseline with recommendations for investigating the research questions around recommendation categories.
\subsection{Procedure}
Sessions lasted approximately one hour, consisting of approximately five minutes of introduction and tutorial, $15$ minutes of closed-ended tasks, $30$ minutes of open-ended exploration, and $10$ minutes of semi-structured interviewing. The tutorial video introduced the interface using the Cars dataset and stated that recommendations were selected based on an interestingness ranking and that the blue text indicated changes from the Current View. For \frontier, the video additionally described each recommendation category. The post-study interview included 7-point Likert scale questions (e.g., overall usability, recommendation usefulness) and open-ended questions on the system design and recommendations. Study scripts and protocols can be found in the supplementary material. 

\subsection{Analysis Approach}
We employed a mixed-methods approach involving both qualitative and quantitative analyses. The primary focus of our work was a qualitative analysis of how recommendations of different categories influenced people's analytical workflows. We conducted a thematic analysis through open-coding of session videos, focusing on strategies participants took to answer their questions. 
\par We thematically classified each participant based on how frequently they engaged with manual controls versus the recommendation panels. To obtain these classifications, we assigned separate labels for characterizing each participant's usage of the Control Panel and the recommendations (1: Majority of the time, 2: Sometimes, 3: Not often). Based on these labels, we grouped the participants by their relative frequency of use, where participants employed a \textit{manual-oriented strategy} if they exhibit a higher usage of the Control Panel than recommendations, \textit{balanced} if they had comparable usage of both, and \textit{recommendation-oriented} if they exhibit a higher usage of recommendations than the Control Panel. Additionally, we define a visualization as \emph{useful} if one or more of the following occurred: (a) the participant verbally described an insight, (b) the visualization was brought into view, (c) the visualization was starred, or (d) the participant expressed that it was useful or interesting. We coded insights from the video recordings, reusing the definition of an insight from prior work~\cite{Sarvghad2017,Liu2014}.
\par The quantitative analysis consisted of Likert question results from the interview as well as counts of expressed hypotheses, data insights, and recommendations participants found useful. We employed statistical testing where appropriate, but considered the quantitative analysis mainly as a complement to our qualitative findings. We adopt a 95\% confidence interval for all statistical analyses. Our analysis approach is similar to other studies that employed mixed-methods to investigate analytical workflows~\cite{mahyar2014supporting, Sarvghad2017}. 

\section{Study Findings\label{sec:analysis}}
\subsection{RQ1: How do recommendation categories support mixed-initiative analysis workflows?}  
To understand how recommendation categories support analytical workflows, we first examine the strategies participants adopted and understand their motivations for switching between different modes of exploration. Then we delve deeper to examine the specific benefits of recommendation categories and their affordances. Finally, we highlight how user's perceptions regarding the recommendation categories can evolve over the course of an analysis workflow.

\subsubsection*{Strategies in mixed-initiative recommendation workflows}
Based on thematic analysis of how frequently participants engaged with manual control versus recommendation panels, we observe three major strategies that they employed across both the \frontier and \baseline conditions. We generally observe that participants were more inclined to use recommendations in their workflow when using \frontier than in the \baseline. We sought to better understand participants' motivations for opting for different analysis options.

Participants employed a recommendation-oriented strategy for exploring unfamiliar attributes ($N_{F,B}=6,3$ participants\footnote{We use the notation $N_{F,B}$ to report measurements for \frontier and \baseline respectively. 
In the example above, $N_{F,B}=6,3$ means that six participants using \frontier and three participants using \baseline used recommendations to explore unfamiliar attributes.}) during preliminary analysis ($N_{F,B}=5,4$) or when they were out of ideas on what to pursue further ($N_{F,B}=5,1$). We found a small group of participants ($N=3$ for \frontier; $N=1$ for \baseline) who relied almost entirely on the recommendations to drive their analyses and used the Control Panel only sparingly.  
Most of these participants either expressed that they had limited experience with creating visualizations or were unsure what to expect from the dataset. The sentiment expressed by these participants largely corresponded to the challenges that visualization novices face in translating abstract questions about their data to visualization specifications~\cite{Grammel2010}. 
As $P6.F$ explained ``\textit{...the recommendations gave me a jumpstart [...] because if I didn't have the recommendations to begin with, I wouldn't even know where to start.}''
\par Participants also adopted a \emph{balanced} strategy intermixing the use of the Control Panel and recommendations in unexpected ways. Three participants ($P7.F$, $P8.F$, $P17.B$) selected recommended visualizations that were ``close enough'' to what they wanted, then made minor tweaks using the Control Panel to attain their desired visualization. Participants also created familiar visualizations to trigger desired recommendations. For example, $P9.F$  wanted to look for linear trends in the data. They recalled seeing a clean linear trend between ACT and SAT scores previously, so they first created the same visualization via the Control Panel. Then they browsed through recommendations resulting from \similarity in order to find similar visualizations. 
Participants were able to leverage recommendations effectively in their workflow since the recommendation categories were transparent and interpretable, leading to predictable behavior.

\par Participants followed a \emph{manual-oriented} strategy when the perceived cost of engaging in manual specification was lower than the effort it took to interact with the recommendations. This  occurs when participants had a specific hypothesis in mind ($P12.F$, $P15.B$, $P16.B$, $P17.B$) or when participants expressed a preference for manual specification due to their familiarity with existing charting interfaces ($P10.F$, $P16.B$, $P22.B$). $P17.B$ explained the reason why they adopted a manual-oriented approach:
	\begin{quote}
	\textit{If the question that I want to answer is very clear, then I will go do it myself.  There are two scenarios that I will switch from the left panel to recommendations. One thing is, I don't know what the next step is and I want insights. Second thing is, I don't know how to do it.}
	\end{quote}
\begin{figure}
	\centering
	\includegraphics[width=\linewidth]{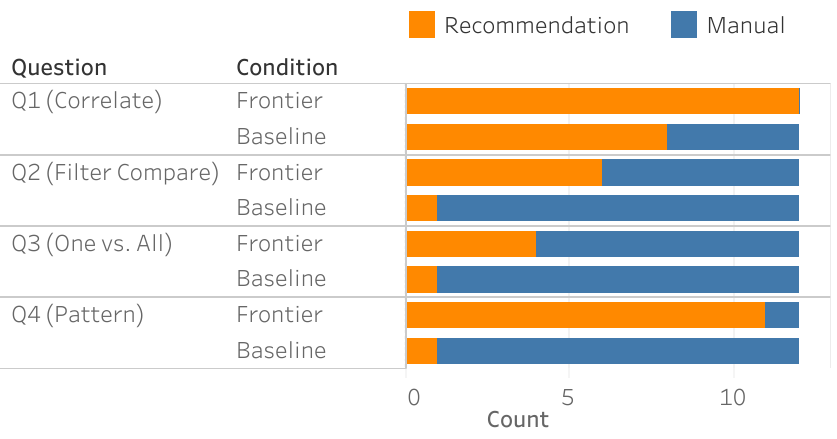}
	\vspace*{-5pt}
	\caption{The number of participants who took a manual-oriented approach in solving the closed-ended question versus a recommendation-oriented approach.}
	\vspace*{-5pt}
	\label{fig:MRstrategy}
\end{figure}
As shown in Figure~\ref{fig:MRstrategy}, a similar pattern is also observed for the strategies taken to solve the closed-ended task. In tasks where manual specification required significantly more work than simply browsing the recommendations for answers (\textsl{Correlate}, \textsl{Filter Compare}), participants were more likely to adopt a recommendation-oriented strategy. On the other hand, in the \textsl{One vs. All} task where participants had to compare a filter and unfiltered visualization, participants opted for the manual-oriented strategy as it was fairly easy to remove a filter. 

Participants also adopted a manual-oriented strategy when the perceived effort to interact with recommendations was higher than usual, such as when they are overwhelmed by the large, unorganized panel of recommendations in the \baseline. This is supported by the post-study Likert ratings, where participants reported recommendations in \baseline to be less useful ($\mu_{F,B}$=$4.58,5.50$; $\sigma_{F,B}$=$0.90,0.79$; \textit{U}=$33.5$, \textit{p}<$0.05$ via Mann-Whitney test) and more overwhelming ($\mu_{F,B}$=$2.58,1.76$; $\sigma_{F,B}$=$1.44,1.76$; \textit{U}=$58.0$, \textit{p}=$0.21$) than \frontier.


\subsubsection*{Value and impact of recommendation categories}
\begin{figure}
	\centering
	\vspace*{-5pt}
	\begin{minipage}{.44\textwidth}
	  	\centering
	  	\includegraphics[width=\linewidth]{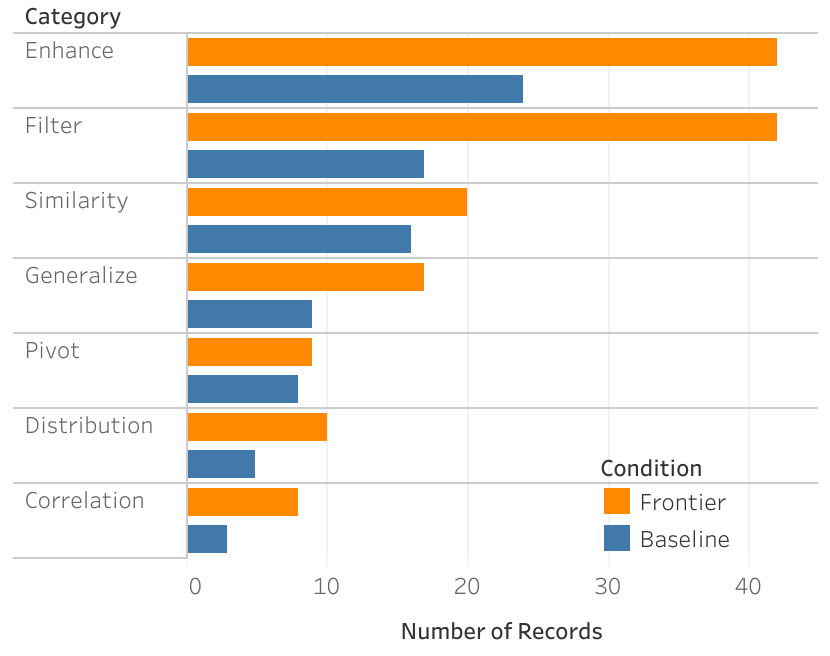}
		\label{fig:UVFacetTypeByCondition}
	\end{minipage}%
	\hspace{20pt}
	\begin{minipage}{.44\textwidth}
		\includegraphics[width=\linewidth]{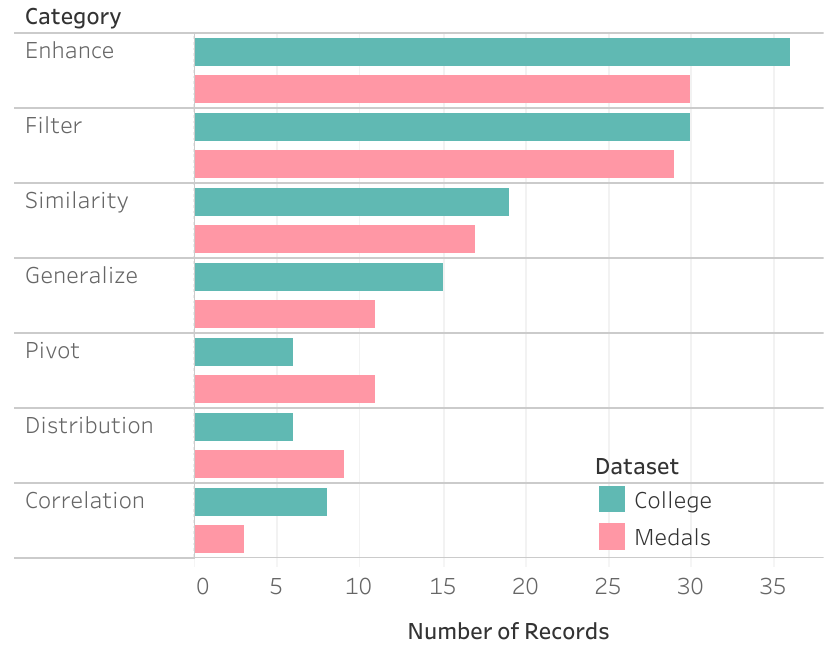}
		\label{fig:UVFacetTypeByDataset}
	\end{minipage}
	\vspace*{-20pt}
	\caption{Number of useful visualizations for each recommendation category by condition (top) and by dataset (bottom) for open-ended tasks. Top: \enhance and \filter are most useful in \frontier, but to a lesser extent in \baseline. Bottom: \pivot and \distribution are more useful in dimension-heavy datasets (e.g., Medals), whereas \correlation is more useful in measure-heavy datasets (e.g., College).}
	\label{fig:UVFacetType}
\end{figure}

We find that the presence of recommendation categories leads to richer and higher-utility exploration. During open-ended exploration, there were more insights generated via recommendations in \frontier than in the \baseline ($N_{F,B}=171,96$; \textit{t}=$2.66$, \textit{p}$<0.05$). A similar trend was observed for the total number of useful visualizations generated via recommendations ($N_{F,B}=149,82$; \textit{t}=2.47, \textit{p}$<0.05$), also shown in Figure~\ref{fig:UVFacetType} (top). 
\par Our observations suggest that recommendation categories reduced overhead associated with interpreting the visualization recommendations. While we did not measure the visualization read-time directly due to the exploratory nature of the tasks, several qualitative observations support this idea. First, six out of $12$ participants using \frontier expressed that they appreciated the organization. $P5$ noted: ``\textit{I like being able to really quickly visually inspect a bunch of things, because I can just slide a bunch of stuff past my eyes and be able to pick the stuff that jumped out.}'' In contrast, many participants in the \baseline condition went back and forth multiple times between visualizations to make comparisons and ensure that they had the right answer when solving the closed-ended questions ($P18$, $15$, $25$, $26$), which, at times, led to mistakes. 
\par During the study, we noticed that some participants appeared to be ``stuck'' in their analyses if they either: a) verbally expressed that they were out of ideas, b) implicitly when they had hypothesizing time of greater than one minute, or c) expressed reluctance to explore further. In particular, only one out of $12$ participants using \frontier got ``stuck'' (once) compared to three participants getting ``stuck'' (total of five occurrences) in the \baseline condition. This is partly attributed to how \frontier participants repurposed and adapted their workflows to take advantage of the diverse set of actions available through various recommendation categories.
\par Participants often leveraged categories with the same axes such as \enhance and \filter to attain insights involving comparisons across multiple visualizations. 
For example, $P10.F$  was interested in the age distribution of Russian athletes because of their highest medal count. They created a histogram distribution of age for Russia and browsed through the \filter action to see distributions for other countries. They exclaimed: ``\textit{Oh wow! Italy has some really old people for their medalists}''. Seeing the Italy age distribution in the context of other age distributions highlighted its uniqueness; the visualization in isolation would have been uninteresting. Such comparisons across visualizations within an axes-consistent category are prevalent and often lead to better distributional awareness and understanding of the general patterns and trends in the dataset. 

\subsubsection*{Evolving perceptions around recommendation categories} 
We found that participants came into the study with a diverse set of perceptions and expectations about recommendations that evolved throughout the course of the open-ended exploration session.
For example, $P8.F$ explained that ``\textit{I feel like these suggestions require a lot more thought process in my head. So for the suggestions, because the one or two times that it doesn't seem a lot useful, I probably disregard it afterwards.}'' $P4.F$ echoed a similar sentiment; several uninteresting visualizations early on deteriorated their confidence in \correlation: ``\textit{I thought that \correlation would be interesting. And it showed me like height and weight correlation and you heard me say I wasn't really interested in that. So it kind of made me nullify the entire \correlation panel together.}'' 
\par We also observed the reverse where participants with a negative initial impression of recommendations gained more trust and understanding over time. $P24.B$ expressed that they had a bias against recommendation systems and was reluctant to look at it. However, finding useful things from the recommendation encouraged them to adopt more of the recommendations in their workflow later on.
    \begin{quote} \textit{Before I even started the study, I have a bias about recommendation panels. Because most of the time recommendation panels do not show you what you want. So I'm already kind of wanting to do my own thing and do it myself, because my bias is that is more reliable than using a recommendation. [...] Once it started showing things, I was like, `Oh, that is kind of interesting', or `Oh, that is kind of relevant'. I started paying a little more attention to it. I had to keep reminding myself like, `Oh yeah, this is the part of the screen I'm supposed to be looking at, it can actually be useful'.}
    \end{quote} 
\par We also observed similar effects on a per-category level. For example, $P11.F$ discovered interesting insights based on \filter and noted in the subsequent analysis that they explicitly focused on the \filter category because they knew it would likely give something interesting. $11$ out of the $24$ participants also expressed that there was a learning curve in familiarizing themselves with the recommendation categories. A more longitudinal follow-up study is required to understand how users would interact with the recommendation categories when they become more familiar. 

\subsection{RQ2: What are the differences in usage and utility across recommendation categories?}
\subsubsection*{Enhance and Filter were most useful, while Pivot least}
As shown in Figure~\ref{fig:UVFacetType}, some categories of recommendations were more useful than others.
Somewhat surprisingly, while participants reported significantly more useful visualizations in \frontier than in \baseline especially for \enhance and \filter, we found that the relative ordering of usefulness for different recommendation types was largely the same independent of the condition. Note that while the categories were not explicitly shown in the \baseline interface, they were logged on the system side for the purpose of this analysis. As shown in Figure~\ref{fig:UVFacetType}, \enhance and \filter are significantly more useful than \pivot (\textit{t}=4.12, p<0.05; \textit{t}=3.24, p<0.05 respectively via t-test).

 
\par In both conditions, we observed that participants manually performed analytical sequences that were similar to the analytical actions that produced the recommendations. We saw repeated patterns of participants manually performing the pivot operation on 14 separate occasions and the filtering operation on seven occasions. Given that \pivot was not actually regarded as very useful compared to the other categories (Figure~\ref{fig:UVFacetType}), we suspect there may be a difference between the types of operation a user would like to perform manually, versus ones that they would prefer being recommended to them. Regardless, these interaction patterns indicate that \filter and \pivot do indeed resemble the natural, intuitive ``next-steps'' in users' workflows. 
\par Participants' post-study ratings of different categories in \frontier largely corresponded to the usefulness counts during the session in Figure~\ref{fig:UVFacetType}. One exception is that \correlation and \distribution were perceived as easily understandable and useful by more than five out of the eight participants who provided a rating, but had low ranks relative to the usage frequencies captured in Figure~\ref{fig:UVFacetType}. The discrepancy likely stems from the fact that these two context-independent categories are only displayed when the Current View is empty, so participants did not see them as frequently as the other categories.\\
  
\subsubsection*{Utility of recommendation categories across \change{datasets}}
As shown in Figure~\ref{fig:UVFacetType} (bottom), the number of useful visualizations from certain categories depended on the dataset. While the usefulness of each category largely followed the trend observed in Figure~\ref{fig:UVFacetType} (left), the usage of \pivot, \distribution, and \correlation differed across datasets. Given that the College dataset contained large numbers of measures, while the Medals datasets had few measures and more dimensions, it was no surprise that there were more uses of \correlation in the College dataset ($N=8$) than in the Medals dataset ($N=3$). 
\par On the other hand, \distribution was used more frequently with the Medals dataset ($N=9$) than in the College dataset ($N=6$), possibly because it also contained bar charts of the count distributions of dimensions (showing a surprising trend that Europe won significantly more medals than any other continent), whereas \distribution for the College dataset showed mostly histograms of measures. \pivot was also used more frequently in the Medals dataset ($N=11$) than in the College dataset ($N=6$), although the we were unable to determine the reason. 

\section{Study Limitations\label{sec:limitations}}
\cut{To avoid information overload, \textit{Frontier} displayed a controlled set of recommendation categories at any point in time based on our category selection logic.} While the analytical workflows that participants chose may be influenced by the category selection logic, we tried to minimize the effect of the display order on category preferences by randomizing the ordering of the various categories across users. We did not investigate the effects of recommendation ranking functions, but instead adopted standard data interestingness metrics from the literature~\cite{datasite,Lee:2019, demiralp2017foresight,Dang:2014,scattersearch}. Future work should explore how the interplay between ranking functions, visualization types, and category labels influences the usefulness of recommendations.

While we employed two different datasets to study task effects, future studies with more realistic dataset properties (large, higher-dimensional), larger sample sizes, different problem domains, and varying user expertise would be helpful. We have not explicitly controlled for visualization expertise, leading to more participants with limited proficiency.
We also acknowledge potential novelty and unfamiliarity effects in our short one-hour study: most participants did not become fully fluent with all the categories and features provided in \textit{Frontier}. The correlation between the `warm-up' closed-ended task and the participant behavior may also be a potential confound. As a result, participants exhibited a strong affinity towards the Control Panel due to preconceptions of and familiarity with existing charting tools. A longitudinal study that examines how categorized recommendations are used in practice is important future work.

\section{Design Implications\label{sec:implication}}
From our study findings, we first describe the guiding principles for the design of \visrec categories. We then discuss interface considerations for recommendation categories and their potential pitfalls. Finally, we identify opportunities for supporting analytical actions in visual analysis.

\subsection{Design guidelines for recommendation categories}
Evidence from our study shows how recommendation categories can be powerful constructs that help situate users by establishing a mental framework for reasoning about recommendation results. The semantic grouping and visual affordances of recommendation categories ``lift'' the visual analysis to operate at the level of analytical actions. By observing how participants switched between  manual specification and recommendations, we find that predictable categories reduce users' perceived cost of employing recommendations --- crucial for establishing an effective mixed-initiative workflow where recommendations can be used seamlessly in conjunction with manual specification.

\par Furthermore, the success of \enhance and \filter lends an important lesson for future \visrec systems in designing simple and readily-accessible recommendation categories. In particular, our study suggests that transparency and interpretability are essential characteristics that lead to recommendation categories that are predictably useful. 

One heuristic for evaluating the complexity of a recommendation is to check whether the underlying action addresses a question involving a single element, which can either be a descriptor of the visualization's characteristics or an element that differs from the current view. For example, \correlation answers the question: ``\textit{Which attributes are correlated?}'' and \enhance answers: ``\textit{What visualizations can be generated by adding one additional attribute?}'' 
Participants' failure to adopt \pivot may be partially due to multiple degrees of freedom in which attributes could be swapped. Further, there may difficulty in articulating the analytical question that \pivot affords. This led to additional cognitive effort to reason about what was retained versus varied. Another contributor might be the drastic encoding change that can occur during swapping. This inconsistent behavior can be jarring and inhibits one's ability to compare across the collection. It remains an open question whether ``anchoring'' techniques that recommends appropriate encodings based \change{on prior context}~\cite{Lin2020} can be applied to recommended collections to offer a more consistent \pivot. Encoding inconsistency across collections is never an issue when swapping out values in \filter as the visualized attributes are unchanged when we move across the value hierarchy. This may explain why \filter was useful in providing complementary views on sub-populations of data, often leading to unexpected insights. 

While our category selection algorithm takes an overview-first~\cite{Shneiderman1996} approach in showing context-independent categories at the beginning and context-dependent recommendations once a specified view exists, several users explicitly cleared their selection in order to get the overview from time-to-time. Additionally, two participants indicated that they hoped to find visualization recommendations that were more unrelated and surprising rather than simple alternatives to their Current View. The diversity-accuracy tradeoff is a classical problem in recommender system design~\cite{Adomavicius}. Supporting a blend of both types of visualization recommendations is a first step towards assisting users with different information preferences and needs. Further research is needed to develop and evaluate these recommendations as well as to validate our proposed taxonomy.

\subsection{Pitfalls of categorized recommendations}
	While recommendations are at most an annoyance when they are not interesting to the user, they can be potentially detrimental if they are used to draw conclusions without deeper examination. $P3.F$  summarized this tradeoff between exploration and exploitation:
	``\textit{For recommendation, sometimes you get completely irrelevant things, things that are kind of normal. But then on the other side, you get this serendipitous discovery, which is very cool. [...]  I mean, it's also dangerous, because you maybe see something where you should further investigate it if it's really an effect.}''
	
Over-reliance on recommendations could be problematic. For example, $P17.B$  said that they had built trust that the system would show something interesting. When the interface did not show anything interesting, they quickly moved onto the next hypothesis instead of digging deeper because they inferred that the system was telling them not to look there. 
We observed a similar effect during a closed-ended task, where participants were asked to select the data sub-populations that had more 8-cylinder cars (Q2). When \frontier displayed only visualizations for three out of the four multiple choice answer options in \filter, many participants who employed recommendations simply drew their conclusions based on the three visualizations included in the category, without verifying the remaining one. 
	\par The potential for erosion of creativity and critical thinking when interacting with an intelligent system is well-known~\cite{Fast2018,datasite}. While this issue is not particular to recommendations based on analytical actions, but a more general phenomenon with recommendations~\cite{Correll2019}, the ease of use and the apparent trust that users perceive from recommendation categories may exacerbate these issues. This challenge points to a need for future research in designing recommendations that provide some notion of coverage or inform users about what has or has not been examined.
\subsection{Towards personalized, adaptive recommendations}
Even though categorized visualization recommendations provide a means of organization, limited screen space makes it impossible to show all categories at once. Furthermore, users typically only peruse the first few items of a recommended list~\cite{Zhou2010}. The diversity of preferences and individual strategies observed in our study suggests that personalized selection of recommendations may be worthwhile For instance, while ten out of the 24 participants believed that there should be fewer recommendations, two participants (one from each condition) wanted to see more.  
\par In post-study interviews, participants indicated that they wanted a more user-driven approach in creating their own organization. Three wanted the ability to extract selected recommendations into a separate dashboard, tab, or page and rearrange them freely into their own groups; eight wanted to hide some or all parts of the recommendation categories and retrieve them on demand. This points to an interesting future direction towards a hybrid human-recommender workflow. Similar to the variability of people's web search behavior~\cite{White:2007}, recommendations could be adaptive and personalized to tailor to users' preferences.
\par Personalization yields potential benefits beyond providing adaptive interfaces, namely, in providing optimization opportunities for system scalability. One of the challenges for visualization recommendations systems is the high computational cost associated with searching over a large search space of possible visualizations~\cite{Vartak2017,Vartak2015,datasite}. Given that there are preferences for certain categories over others for different users, datasets, and tasks, there is an opportunity to reduce the computational cost of an exhaustive search by pruning the recommendation search space. 

\section{Beyond The Final Frontier}
The goal of recommender systems is in some sense to anticipate future user needs. Recommendation categories help organize these possible futures as readily-available options to drive analytical workflows. We introduced a taxonomy based on prior literature to examine the usefulness of various recommendation categories based on the underlying analytical actions. We implemented \frontier as a design probe to better understand how users push towards the frontier, taking next steps in their exploration. Our user study confirmed that recommendation categories are indeed useful for facilitating data exploration, helping users understand and interpret the visualizations. While the general utility of categorization was not surprising, we more deeply explored \emph{how} various categories of visualization recommendations were employed and the diverse workflow strategies that users adopted. Design implications stemming from this study provide unique opportunities for supporting delightful user experiences with next-generation \visrec systems.



\bibliographystyle{IEEEtran}
\bibliography{frontierbib}

\end{document}